\documentclass[onecolumn,amsmath,noshowkeys,noshowpacs,amssymb]{revtex4}

\usepackage{graphicx,color}

\usepackage{amsbsy}
\usepackage{amsthm,mathrsfs,amsopn}
\usepackage{mathtools}
\usepackage{cases}
\usepackage{float}
\usepackage{bbm}

\newcommand{\rvline}{\hspace*{-\arraycolsep}\vline\hspace*{-\arraycolsep}}

\begin{document}

\title{Recurrent Spectral Network (RSN): shaping the basin of attraction of a discrete map to reach automated classification}

\author{Lorenzo Chicchi$^1$, Duccio Fanelli$^1$, Lorenzo Giambagli$^{1,2}$, Lorenzo Buffoni$^{1,3}$, Timoteo Carletti$^{2}$\vspace*{.25cm}}

\affiliation{$^1$ Dipartimento di Fisica e Astronomia, Universit\`a di Firenze, INFN and CSDC, Via Sansone 1, 50019 Sesto Fiorentino, Firenze, Italy}
\affiliation{$^2$ naXys, Namur Center for Complex Systems, University of Namur, rue Graf\'e 2, B 5000 Namur, Belgium}
\affiliation{$^3$ Physics of Information and Quantum Technologies Group, Instituto de Telecomunicacoes, University of Lisbon, Portugal}

\begin{abstract} 
A novel strategy to automated classification is introduced which exploits a fully trained dynamical system to steer items belonging to different categories toward distinct asymptotic attractors. These latter are incorporated into the model by taking advantage of the spectral decomposition of the operator that rules the linear evolution across the processing network. Non-linear terms act for a transient and allow to disentangle the data supplied as initial condition to the discrete dynamical system, shaping the boundaries of different attractors. The network can be equipped with several memory kernels which can be sequentially activated for serial datasets handling. Our novel approach to classification, that we here term Recurrent Spectral Network (RSN), is successfully challenged against a simple test-bed model, created for illustrative purposes, as well as a standard dataset for image processing training.
\end{abstract}


\maketitle

\vspace{0.8cm}

\section{Introduction}
\label{sec:intro}

Machine learning (ML) technologies are becoming increasingly popular due to their inherent degree of transversal adaptability, which transcends different realms of applications \cite{he2018amc, sutton2018reinforcement, grigorescu2020survey, biancalani2021deep, Goodfellow-et-al-2016, lecun2015deep,deng2014deep}. Multi-layered feedforward neural networks, notably the most basic architecture schemes, are routinely employed to cope with a large plethora of case studies and test-bed models. ML seeks at solving an optimization problem, upon minimization of a suitably defined loss function which confronts the expected target to the output produced at the exit layer, after a nested sequence of linear (across layers) and non-linear (punctually localized on the nodes) manipulations of the data supplied as an entry.  The target of the optimization are the weights of the links that connect pair of nodes belonging to adjacent stacks of the multi-layered arrangement, in a fully coupled setting. An alternative training scheme has been recently proposed which anchors the learning to reciprocal domain \cite{spec_learn,chicchi2021training}: the eigenvalues and the eigenvectors of the transfer operators get adjusted by the optimization. Spectral learning, so far engineered to deal with a feedforward organization, identifies key collective variables, the eigenvalues, which are more fundamental than any other (randomly selected) set of identical cardinality, allocated in direct space. Extending the learning to include the eigenvectors enhances the ability of the network to carry out the assigned task. 

Delving into the principles of the spectral methodology, we here propose a radically novel approach to computational machine learning which is deep rooted into the theory of discrete dynamical systems. In a nutshell, the incoming signal is processed by successive iterations across the very same constellation of nodes. The links, and thus the topology of the ensuing network, are fixed and shaped under the spectral paradigm, upon optimization at a given number of iterations. Non-linearities acting on the nodes are imposed a priori or, conversely, learned self-consistently via an apposite deep neural network, which is embedded into the cost function. In either settings, non-linear terms are forced to vanish asymptotically, iteration after iteration, in such a way that the dynamics eventually turns purely linear. The linear operator, mirroring the processing network, is enforced with a set of distinct attractors, namely the eigenvectors associated to the eigenvalues equal to one, each pointing to a specific class of items to be at last categorized. Stated differently, the classification is accomplished when the processed output aligns along a specific direction in dual space, instead of turning active a single node in direct space, as customarily done. This formulation yields a rather natural interpretation of the classifier operational mode: non-linearities, acting at the early stages of the dynamical evolution, shape the basin of attractions of the imposed stationary equilibria. Different elements, read at the entry and interpreted as the initial condition of the trained dynamical system, fall in distinct basins of attractions according to their specificity. Delineating the non-trivial contours that entrap the basins of attraction constitutes the tangible outcome of the learning scheme. Remarkably, the trained dynamical system can be iterated forward in time, beyond the limited horizon of the learning procedure: the ability of classifying stays unchanged.  The eigenvectors associated to eigenvalues equal to one, are veritable memory kernels where the information is kept stored. We name Recurrent Spectral Network (RSN) our novel approach to automated classification via sculpting the basin of attraction of a discrete dynamical map.

Points of connections are found with the framework of reservoir computing. In this latter case, input signals are mapped into higher dimensional computational spaces through the dynamics of a fixed, non-linear system termed reservoir \cite{gauthier2021next, tanaka2019recent, maass2002real}. Within the RSN, the bulk model is not fixed but self-consistently tailored to the assigned task.

A straightforward variant of the RSN recipe, which accounts for quasi-orthogonal eigen-directions for each processed task, can be also introduced. This latter enables for the sequential handling of different datasets. In simple terms, an artificial computing unit can be assembled which keeps memory of a task, for which it was initially trained, while being exposed to another training session, with an independent dataset to be processed. This is at present arduous with standard approaches to machine learning, as the second learning stage causes the so-called catastrophic forgetting taking over any form of digital consciousness inherited from the first \cite{mccloskey1989catastrophic,lewandowsky1995catastrophic,kemker2018measuring}. Few attempts have been so far reported which aim at overcoming this limitation \cite{kirkpatrick2017overcoming,goodfellow2013empirical,li2019learn}. 

The paper is organized as follows. In the next Section we intoduce the mathematical notation and the relevant model setting. Then, in the subsequent Section, we will turn to considering a simple example of a dataset defined in $\mathbb{R}^2$ that will prove useful for clarifying the essence of the proposed methodology. In particular we will show, that the system can effectively trace the boundaries that non-linearly separate different classes within a given datasets, each of them being directed toward a distinct asymptotic attractors. Further, we will proceed by applying the proposed technique to the celebrated MNIST dataset \cite{lecun1998mnist}. We will also show that the RSN can also handle multiple datasets with a modest drop in the peak accuracy,  and following sequential stages of learning. Finally, we will sum up and draw our conclusions. 

\section{The mathematical foundation}

Consider $N$ isolated nodes. Our aim is to assign weighted links among the latter, in such a way that the ensuing network can cope with the assigned task, 
as e.g., classification of different items in distinct categories.  Here, $N$ can coincide with the number of input variables (e.g., the pixels of a supplied image): in this case, the nodes where  reading is performed match the units where calculations are carried out. This is at variance with usual feedforward deep neural networks, where the information to be processed flows from the input to the output, the collection of computing neurons growing with the number of layers that define the underlying architecture \cite{lecun2015deep,Goodfellow-et-al-2016,deng2014deep}. Working within the proposed framework, the topology of the network will unfold as an emerging byproduct of the optimization procedure.  As we shall discuss, $N$ can be larger than the characteristic dimension of the input data, a setting that we will specifically assume when dealing with the problem of sequential learning, with dedicated memory kernels. 

Denote by $\vec{x}^{(0)}$ the input vector, made of $N$ entries organized in a column. The idea that we shall hereafter develop is to set up a recursive scheme, the Recurrent Spectral Network (RSN), that takes $\vec{x}^{(0)}$ as the initial condition and transforms it via successive iterations into a stationary stable output. This latter should somehow reflect the specific traits of the input items, as identified self-consistently upon dedicated training sessions. Different objects should be directed towards distinct  attractors, depending on the category of specific pertinence. Stated differently, the multidimensional space where the examined objects belong to gets partitioned in mutually exclusive basins of attraction, as  tailored by suited non-linearities, each associated to a definite asymptotic destination.  In the following, we shall  label with $n$ the number of independent attractors, namely the number of independent classes in which the inspected dataset can be eventually partitioned.

Assume $\vec{x}^{(k)}$ to represent the image of the input vector $\vec{x}^{(1)}$ after $k$ application of the iterative scheme. Then:

\begin{equation}
\vec{x}^{(k+1)}= \vec{f}_k \left( {\mathbf A} \vec{x}^{(k)} \right)
\label{nlinear}
\end{equation}

where ${\mathbf A}$ is a $N \times N$ weighed adjacency matrix that defines the patterns of interactions among nodes; $\vec{f}_k(\cdot)$ is a non-linear ($N$- dimensional) function that depends on the iteration parameter $k$ and which acts at the level of individual  nodes. We require in particular $\lim_{k \rightarrow \infty} \vec{f}_k \rightarrow \vec{\mathbbm{1}}  \equiv \left(1, 1, ..., 1\right)^T$, in such a way that, for large enough $k$, the system approximately follows a linear update rule. This is achieved by setting:

\begin{equation}
\vec{f}_k(\cdot) = \vec{\mathbbm{1}} + \frac{\vec{g}(\cdot)}{k^{\gamma}}
\end{equation}

where $\vec{g}(\cdot)$ is a non-linear function which can be imposed a priori or determined self-consistently via a neural network regression model and $\gamma$ is a parameter that can be freely adjusted (here we chose to set $\gamma=1.5$). Focus now on the linear component of the dynamics, as encapsulated in matrix ${\mathbf A}$, which takes over for sufficiently large $k$. We cast in particular ${\mathbf A}={\mathbf \Phi} {\mathbf \Lambda} \left({\mathbf \Phi}\right)^{-1}$, by invoking spectral decomposition. Here , ${\mathbf \Lambda}$ is the diagonal matrix of the eigenvalues $(\lambda_1,\lambda_2, ..., \lambda_N)$.  Working in the spectral domain enables us to enforce the sought attractors, which totals in $n$.  To this end, we impose $\lambda_1=\lambda_2=...=\lambda_n=1$, and assume  $|\lambda_i|<1$ for $i>n$. These latter $N-n$ quantities are among the target of the optimization scheme. Moreover, we assume $\vec{\phi}_1$, $\vec{\phi}_2$,..., $\vec{\phi}_n$, namely the eigenvectors relative to the eigenvalues identically equal to one, to identify frozen linearly independent directions of the embedding $N$-dimensional space.
The remaining eigenvectors ($\vec{\phi}_i$, with $i>n$,  relative to eigenvalues  $\lambda_i$) can be freely adjusted, so contributing with a total of $(N-n) \times N$ tunable parameter to the optimization scheme. When $k >> 1$,  non-linear terms fade away and the iterative scheme converges to a linear map, $\vec{x}^{(k+1)} \simeq {\mathbf A} \vec{x}^{(k)}$. By definition,  $\vec{\phi}_i$, with $i \le n$ are stationary solutions of the above linear equation, which hence possesses $n$ a priori shaped attractors (their linear combinations are also stationary solutions of the dynamics). The same holds true for the non-linear counterpart model (\ref{nlinear}), which is by construction designed to approach its linear limit for large enough iterations $k$. By acting on the collection of tunable spectral parameters, which ultimately echo on the topology of the network made of $N$ computing nodes, and exploiting the non-linearities that act over a finite transient, we aim at steering different input objects toward distinct target attractors, which can be stably maintained beyond the limited horizon of the performed training.  To rephrase in words, we postulate that any generically complex classification task si eventually amenable to a multi-dimensional linear problem, with properly tuned interactions strengths and provided non-linearities, imposed or self-consistently learned,  are made to initially deform the features landscape.

To implement the learning scheme on these basis, we consider $\vec{x}^{(\bar{k})}$, the image on the output layer of the input vector $\vec{x}^{(0)}$ after $\bar{k}$ iterations of the iterative algorithm, where $\bar{k}$ is sufficiently large for the linear approximation to hold true. Then, we calculate $\vec{c}_{\bar{k}}=\left({\mathbf \Phi}\right)^{-1} \vec{x}^{(\bar{k})}$: the $i$-th element $\left(\vec{c}_{\bar{k}} \right)_i$ represents the projection of $\vec{x}^{(\bar{k})}$ along the 
 eigen-direction $\vec{\phi}_i$. Each element of the training set is associated to a label $\ell \le n$ to identify the category to which $\vec{x}^{(0)}$ belongs to. Then,
the optimization scheme is simply set which seeks at minimizing the squared distance of  $\vec{c}_{\bar{k}}$ (which implicitly depend on the training parameters) with a target $n$-dimensional column vector $\vec{c}_{\ell}$, which is made by zeroes except for the element in position $\ell$ which is set to unit.

In such a way, we require that after sufficiently many iterations the dynamical map aligns along the direction  $\vec{\phi}_{\ell}$, where we recall  that $\ell$ identifies the class to which the supplied entry refers. Different initial conditions, decorated with their reference labels pointing to one of the $n$  classes, are forced (by a proper use of the non-linearities, as vehiculated by the network arrangement) to fall in separated basin of attractions of the trained dynamical model, each yielding a different asymptotic stationary stable equilibrium, e.g., a direction in reciprocal space. In the following Section, to challenge the effectiveness of the proposed recipe, we set to study a simple yet non-linearly separable dataset defined in $\mathbb{R}^2$, which bears pedagogical interest. We will then turn, in a subsequent Section, to examining the ability of the RSN methodologies to cope with a standard datasets of image.

\section{Testing RSN:  a simple dataset in $\mathbb{R}^2$}

As mentioned above, we aim at testing the RSN as outlined above against a simple dataset, created for this specific purpose. The goals  are twofolds. On the one side, we wish to provide the first consistent implementation of the procedure, by showing that a dynamical system can be trained which preserves its ability to discern beyond the horizon of the training (as instead it is the case for conventional recurrent neural networks). This is an indirect mark of the imposed convergence towards an asymptotic equilibrium, inherent to the dynamical scheme, which flags the class to be identified. Then, we shall convincingly demonstrate that classification by RSN amounts to segmenting the space of the initial conditions in disconnected domains, the basin of attraction of the implanted attractors.

\begin{figure}[h]
	\centering
	\includegraphics[width = 0.4\textwidth]{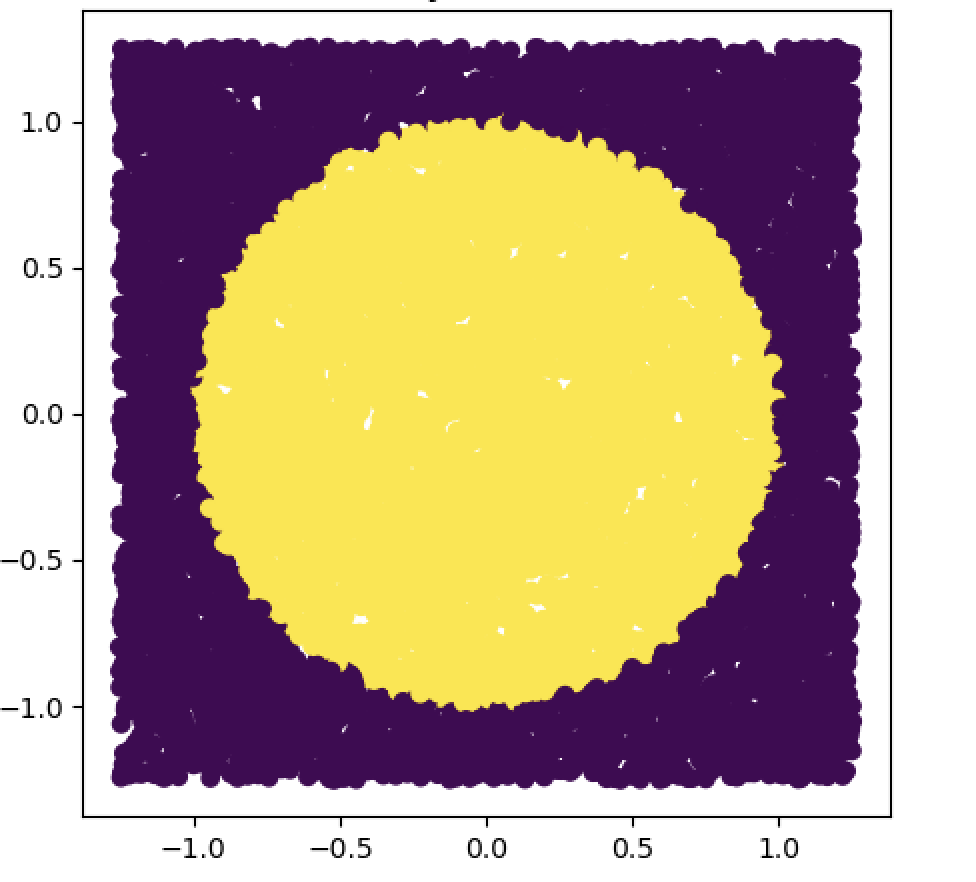}
	\caption{The dataset used as a validation test for the RSN scheme. Points populate two different regions, of equal relevance, separated by a sharp non-linear boundary, which we identify as the unitary circle.}
	\label{f:easy_dataset}
\end{figure}

The dataset that we shall here consider as a proof of concept is composed by two sets of points, laying on the plane. The points falling inside the unitary circle, centred at the origin, define the first class (displayed in yellow, in Figure \ref{f:easy_dataset}). Those situated outside the circle and inside a square domain of linear width $L= \sqrt{2 \pi}$, contribute to the second reservoir of datapoints (shown in blue, in Figure \ref{f:easy_dataset}). The size of the square has been chosen in such a way that the surface of the two regions where the dataset insists is equal. The two sets are divided by a non-linear boundary that coincides with the perimeter of the unitary circle. It is hence not possible to distinguish among the two datasets by means of a linear classifier. Our objective is to train a RSN, following the prescriptions of the preceding Section, so as to associate  any given point - randomly generated to belong to the square domain of width $L$ - to its reference portion, as introduced above.  

For the sake of definiteness we cast $N=10$. Every point of coordinates  $(x,y)$ (constrained so as to fall inside the square of linear size $L$) yields an initial condition for the RSN that we wish at training, i.e. $(\vec{x}^{(1)})=(x,y,0,0,0,0,0,0,0,0)$. During the training stage, we generate a sufficiently large reservoir of ($M$) points, each complemented with a scalar label that specifies the class, or domain, where the corresponding point falls. The first two eigenvalues of 
${\mathbf \Lambda}$ are set to unit and the corresponding eigenvectors, respectively $\vec{\phi}_1$ and $\vec{\phi}_2$, are fixed and identify randomly selected (linearly independent) directions in $\mathbb{R}^{10}$. The eigenvalues $\lambda_i$, as well and the entries of the vectors $\vec{\phi}_i$, for $i>2$, contribute to the pool of parameters that  one can freely adjust during optimization. Moreover, and to test the method in its general formulation, we represent the non-linear function $g(\cdot)$ (the very same function for each node of the RSN) as a two layered neural network. Each of these latter layers is  made of $30$ neurons and nodes are entitled with a $\tanh$ activation function. We label with $\bar{k}$ the number of iterations of the RSN, assumed during training. Recall that we will be also interested in assessing the behavior of the fully trained systems for $k>\bar{k}$. In the following 
$\bar{k}=60$. The number of epochs is set to 200 and an early stopping technique has been employed.

\begin{figure}[h]
	\centering
	\includegraphics[width = 0.5\textwidth]{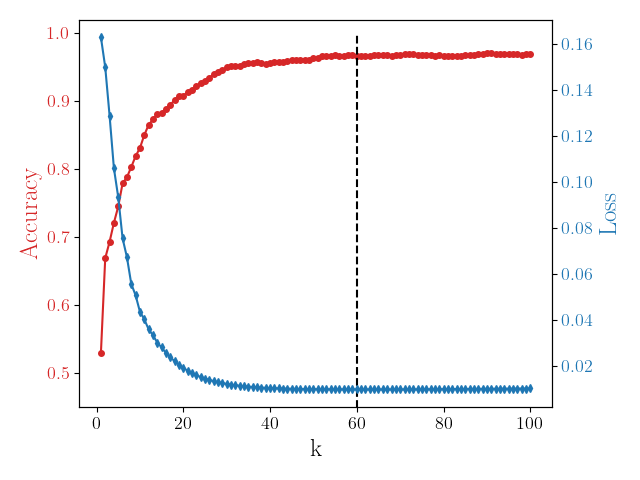}
	\caption{Accuracy (in blue) and loss (in red)  against the iteration $k$, for a trained RSN with $\bar{k}=60$ (vertical dashed line). Data refer to just one realization of the training procedure.}
	\label{f:acc_loss_easy}
\end{figure}

In Figure \ref{f:acc_loss_easy}, the  test-accuracy and the corresponding loss are plotted for $ k<\bar{k} $ and for $ \bar{k}<k<100 $. As it can be visually appreciated, the accuracy (and the loss) is stable for $k>\bar{k}$, i.e., when extending the RSN beyond the iteration number assumed for training.

The trained RSN classifies points  $(x,y) \in \mathbb{R}^2$, provided as an input, by generating a late time output in $\mathbb{R}^{10}$ which tentatively aligns along different target directions: points in the plane contained within the unitary circle with center in the origin, should predominantly activate the spectral mode $\vec{\phi}_1$.  In this case, $c_1$ is thus expected to stand out, as compared to all others coefficients, after sufficiently many iterations. At variance, points falling outside the unitary circle are dynamically driven towards a final equilibrium which selectively favours the eigen-direction $\vec{\phi}_2$. The coefficient $c_2$ should therefore prevail over the others. This scenario is confirmed by inspection of Figure \ref{f:ceff_vel_easy}, where $c_1$ and $c_2$ are plotted against the iteration number for data points falling respectively inside (top panel) and outside (lower panel)  the  unitary circle. Different classes are hence associated to distinct attractors, as stipulated a priori. Indeed, any direction obtained as a linear combination of $\vec{\phi}_i$  with $i=1,2$, is also, by construction, a stationary solution of the RSN. This is why a residual activation of the other modes - those relative to eigenvalues one but different from that identified as the target for the class under scrutiny - can in principle manifest when the  RSN is challenged against the test-set. A softmax mask, applied after the final iteration in dual space, would enforce a perfect alignment along the sought target direction, with no impact on the performance of the trained device. 

\begin{figure}[h]
	\centering
	\includegraphics[width = 0.5\textwidth]{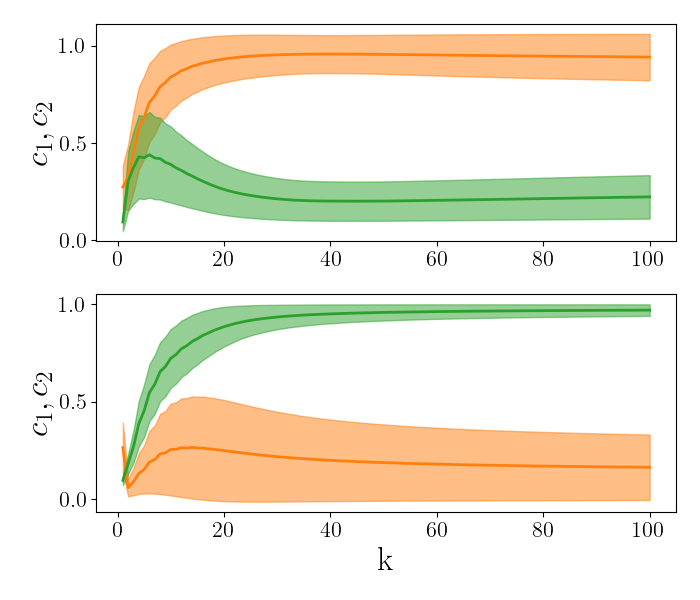}
	\caption{The evolution of the coefficients $c_1$ (orange) and $c_2$ (green) is plotted for points of the test set positioned respectively inside (top panel) and outside (lower panel)  the  unitary circle. The shadowed region points to the standard deviation of the collected signal when averaging over the population of supplied input, organized in groups which reflect their domain of pertinence.}
	\label{f:ceff_vel_easy}
\end{figure}

\begin{figure}[h]
	\centering
	\includegraphics[width = 0.5\textwidth]{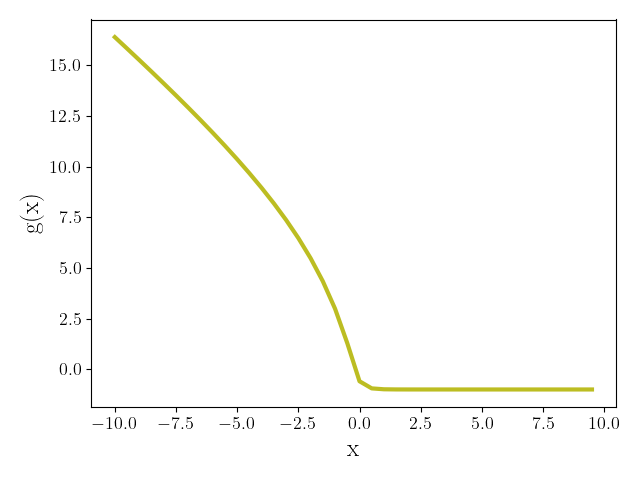}
	\caption{The non-linear function $g(x)$ as obtained from the regression neural model that is associated to each computing neuron of the RSN.}
	\label{f:nonLin}
\end{figure}

The above analysis carried our for a simple benchmark model allowed us to grasp some intuition on the decision making scheme as implemented via the dynamical RSN. Classification is here synonym of convergence towards a specific attractor, which is flagged as the destination target of the dynamics, for a homogeneous ensemble of input items.  Different classes are hence interpreted by the RSN as the basins of attractions of the stationary modes, the eigen-directions of ${\mathbf A}$ associated to eigenvalues equal to one. For the case at hand, the separatrix between the domains in  $\mathbb{R}^2$ which defines the two classes to be eventually identified matches the unitary circle. To show that the RSN is able to correctly spot out the non-linear separation between the two contiguous domain in $\mathbb{R}^2$, and so resolve the distinctive features of the dataset under exam, we consider $\langle c_i \rangle$, the average of the $i$-th coefficient, across successive phases of the RSN evolution and for different input choices $(x,y) \in \mathbb{R}^2$. More specifically,  $\langle  c_i \rangle (x,y)= \frac{1}{k_F-k_I}\sum_{k = k_I}^{k_F} (c_k)_i(x,y)$, for all specific coefficients $i$ - including those which will fade away after a transient - and as function of the departure point.  In Figure \ref{f:ceff_vel_easy} the computed coefficients are displayed in the reference plane $(x,y)$, with an apposite colorcode and for different choices of $(k_I, k_F)$. The panels on the top refers to the initial stages of the evolution ($k_F=5, k_I=1$): the separation between the two classes here considered leaves a clear imprint in the distribution of the $\langle c_i \rangle$ (in particular those with $i>2$) across $(x,y)$ (in Figure \ref{f:ceff_vel_easy} we plot $\langle c_6 \rangle$, as an illustrative example as well as $\overline{\langle c \rangle }=\left( \sum_{i=3}^{10} \langle c_i \rangle \right)/7$).  An abrupt transition is indeed observed for $\langle c_i \rangle$, with $i>2$, when crossing the unitary circle, namely the separatrix between the two adjacent classes that defines our test model. The patterns associated to $\langle c_1 \rangle$ and $\langle c_2 \rangle$ are less clear, at short times, but become evidently distinct when  the iterations number is made to increase (see lower panels, referred to $k_F=50$, $k_I=40$). Transient modes (those associated to eigenvalues with magnitude smaller than unit)  are employed for an early assessment of the examined dataset and get progressively disangaged, at later times. The processed information is in fact passed over the stationary directions, where it is eventually crystallized for classification purposes. Averaged projection coefficients can be employed to trace out, in direct space,  key distinctive features that form the basis of decision making (the boundaries of distinct basins of attractions).  It is here speculated that this is a general attribute of the RSN that can be exported to other, more complex, settings for an a posteriori understanding of the principles that guide artificial reasoning. As a side complement, in Figure \ref{f:nonLin} we depict the non-linear function $g(\cdot)$  self-consistently obtained via the regression neural model accommodated for in the RSN. In this specific case, it looks like an inverted ReLu  with an additional offset.

\begin{figure}[h]
	\centering
	\includegraphics[width = 0.7\textwidth]{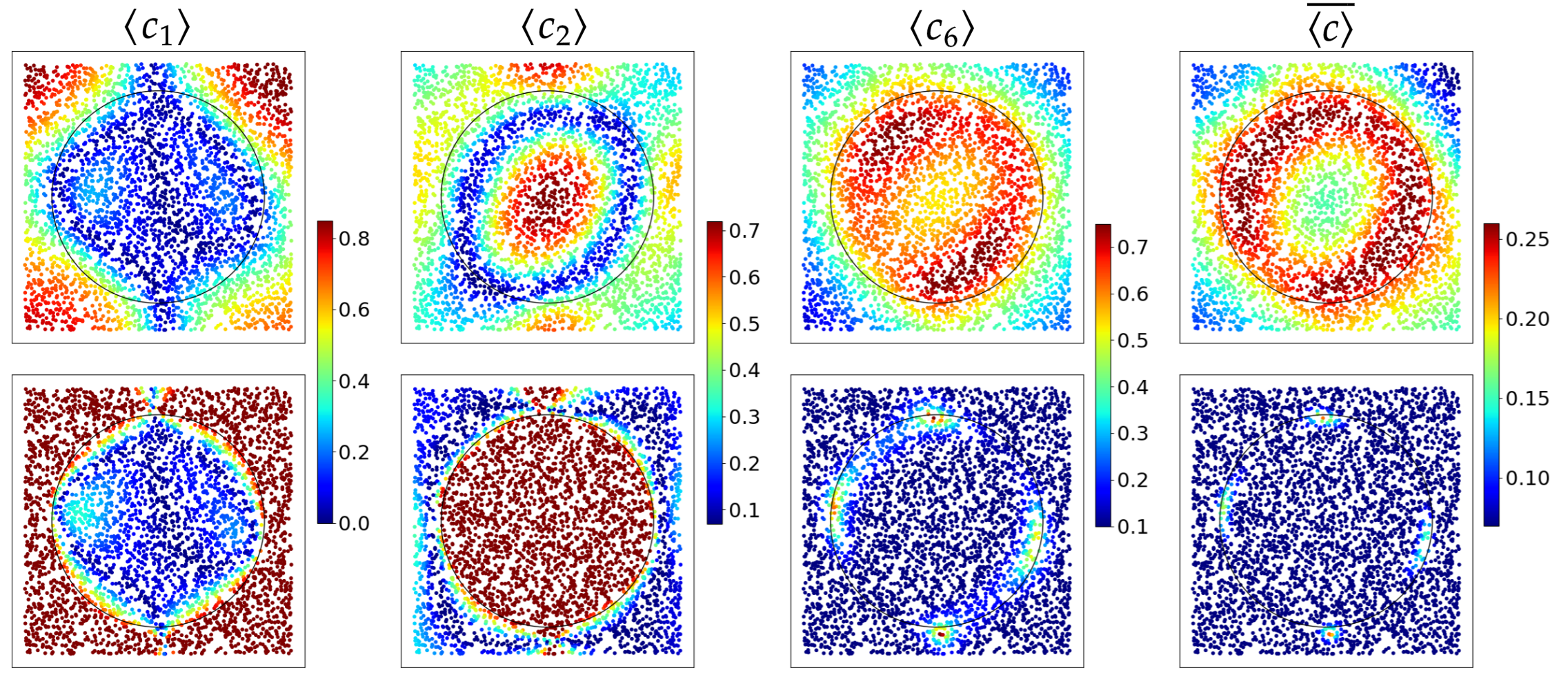}
	\caption{The quantities $\langle c_i \rangle$ for $i=1,2,6$ and $\overline{\langle c \rangle}$ are plotted, for different $(x,y)$, i.e moving on the plane of the initial condition. Top panels refer to $k_F=5, k_I=1$. Lower panel to $k_F=50, k_I=40$. The separatrix between the two considered classes  
	(which coincides with the unitary circle centered at the origin) is sensed, at short times, by the transient directions. The projections of the generated output along these latter directions fade asymptotically away and the existence of the two classes, as well as the relative domain of definition, leave an imperishable trace in $\langle c_1 \rangle$ and $\langle c_2 \rangle$.}
	\label{f:ceff_vel_easy}
\end{figure}

Building on these preliminary observations, we will turn in the next Section to considering the application of RSN to MNIST dataset.  

\section{Applying RSN to the MNIST dataset}

As a further step in the analysis, we apply the RSN to the celebrated  MNIST dataset \cite{lecun1998mnist}. This is a collection of handwritten digits: the training set consists of 60,000 examples, and a test set of 10,000 examples. Each image is made of $N=28 \times 28=784$ pixels and each pixel bears an 8-bit numerical intensity value. The images are to be classified in 10 distinct groups (the numbers from $0$ to $9$). Each element of the training set is associated with an integer label to point to the class to which the selected image belongs to. In the following we will set to train a RSN made by $N=784$ nodes: the nodes that receive the information as an input are the very same nodes that carry out the classification, through a dynamical segmentation that originates from the underlying RSN. The network of excitatory (positive weight) or inhibitory (negative weight) interactions is shaped by the optimization scheme which seeks at adjusting the non trivial eigenvalues and eigenvectors of matrix ${\mathbf \Phi}$. The first $10$ eigenvalues are set to unit, as in the spirit of the above, and refer to the eigen-directions employed for discrimination. These latter eigenvectors are a priori fixed and can be engineered so as to return evocative patterns in the space of the inspected images, as we shall demonstrate in the following. Further, we assume $g(\cdot)=\tanh(\cdot)$, for the sake of simplicity. Summing up, we can count on a total of $N \times (N-10)+(N-10)$ adjustable parameters to yield a fully trained RSN which can efficiently classify MNIST images. 

\begin{figure}[H]
	\centering
	\includegraphics[width = 0.5\textwidth]{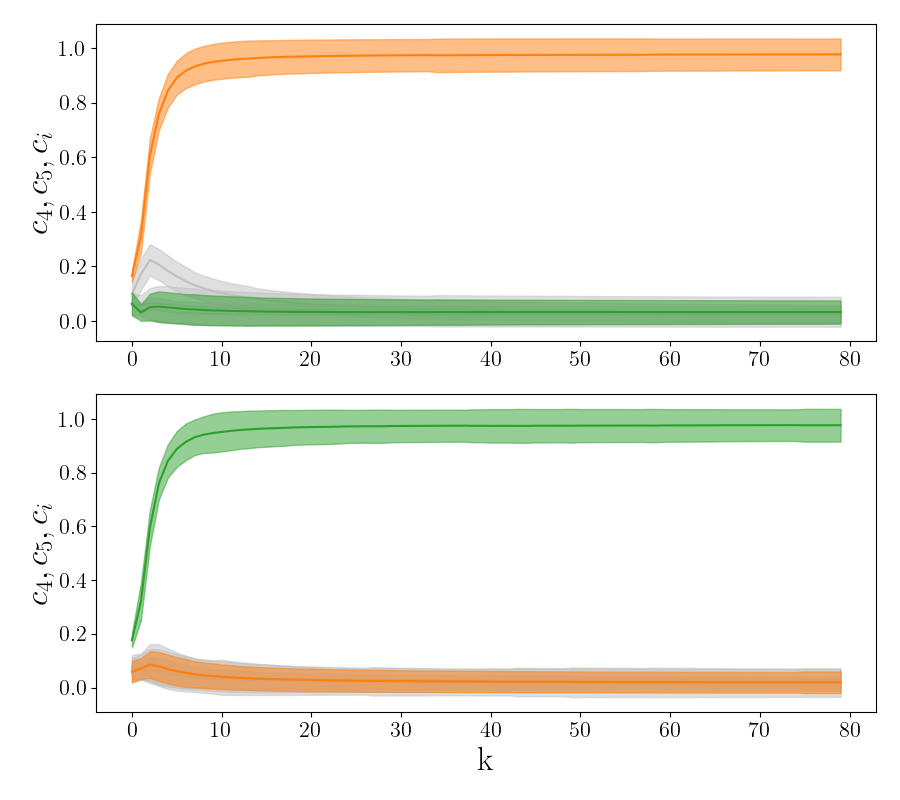}
	\caption{Top panel: the full set of handwritten four available in the test set is provided as an input to the trained RSN and the response monitored in terms of the obtained $c_i$, with $i=1,...,10$. As expected, $c_4$ (orange) emerges and converges to unit, for $k>10$ ($\bar{k}=10$ being the maximum iteration number set during training). All other coefficients, including $c_5$ (green) disappear. Lower panel: the situation is analogous to that analyzed in the top panel with the notable exception that now handwritten five are analyzed by the RSN. Hence, $c_5$ (green) converges to unit while, $c_i$ with $i \ne 5$ (including $c_4$, in orange) fade away. In  both cases, the shadowed regions reflect the variability of the images, within any given class of the test set.}
	\label{f:convergenceMNIST}
\end{figure}

In Figure \ref{f:convergenceMNIST}, we challenge the ability of the trained RSN to discern images of the test set that respectively corresponds to four (top panel) and five (lower panel). In the former case, as expected,  $c_4$ (depicted in orange) sticks out as the only residual coefficients after sufficiently many iterations of the RSN machinery. All other coefficients (including $c_5$, plotted in green) are eventually bound disappear, thus implying that the reservoir of images supplied as an input are eventually directed toward a unique attractor. Stated differently,  the collection of handwritten four is being seen by the RSN as belonging to a specific basin of attraction. The shadowed regions that are associated to each average curve refer to the degree of variability inherent to the examined gallery of images. The lower plot in Figure \ref{f:convergenceMNIST} shows the response of the RSN when the images displaying a number five are read as an input, and the interpretation is in  line with the above. In both cases, the training is performed by arresting the RSN at iteration $\bar{k}=10$: the outcome is however stably maintained well beyond the training horizon, with a modest, although significative in terms of its philosophical implications, improvements in terms of confidence of the assessment. When it comes to the overall performance, the accuracy on the train set is of about $98 \%$, while on the test set the RSN scores $97 \%$, in line with what usually reported when using conventional approaches to machine learning.

\begin{figure}[H]
	\centering
	\includegraphics[width = 0.4\textwidth]{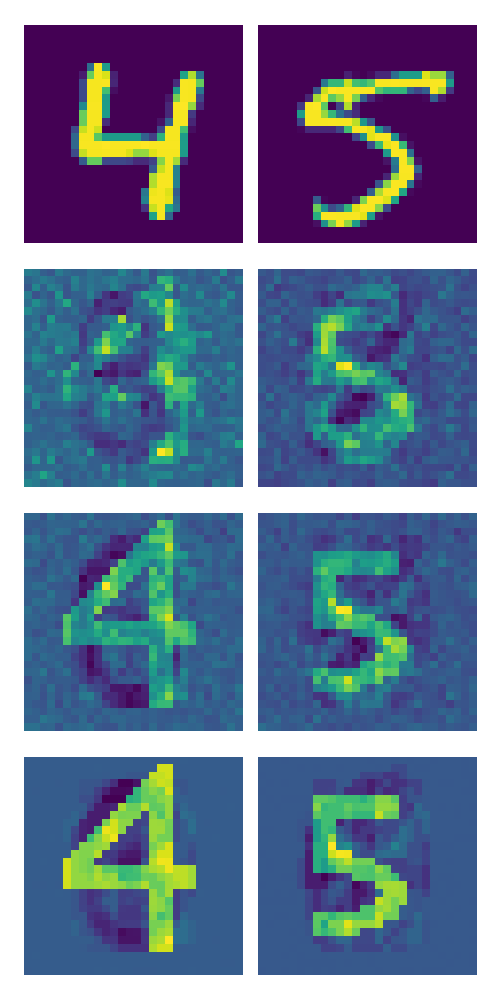}
	\caption{In each row we plot the activity on each node of the RSN, at different iterations and for two input numbers that belong to two distinct categories, respectively a four (left) and a five (right), see top panels. After a few iterations the RSN converges asymptotically to the attractors (respectively the eigenvectors $\vec{\phi}_4$ (left) and $\vec{\phi}_5$ (right))  that are triggered by the provided input. Notes that the attractors can be shaped to manifest as a stylized version of the number to be classified. The more yellow the pixels, the more intense the activity on the associated nodes.}
	\label{f:conver45}
\end{figure}

Figure \ref{f:conver45} illustrates the progressive convergence of the scheme, for two distinct exemplaries of input images. The RSN converges asymptotically to the deputed attractors, which respectively correspond to eigenvectors $\vec{\phi}_4$ (left) and $\vec{\phi}_5$ (right). The entries of these latter eigenvectors are shaped so as to return a stylised version of the digits that define the categories in which the dataset is partitioned. The outcome of the analysis is hence a stationary stable image, the plastic modulation  of the input that is dynamically steered towards a final destination shaped at will by the operator.

\begin{figure}[H]
	\centering
	\includegraphics[width = 0.5\textwidth]{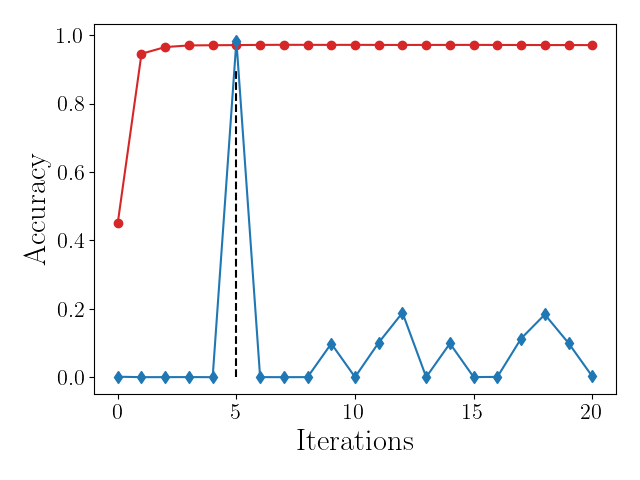}
	\caption{Evolution of the accuracy as computed on the test set of the MNIST dataset. Red symbols stand for a RSN model, trained at $\bar{k}=5$ (black dashed line); blue symbols refer to a RNN, with $\bar{k}+1$ consecutive layers, i.e. with $\bar{k}$ nested applications of the same $N \times N$ transfer operator. The RSN quickly converges to the best accuracy, which stays constant for $k>\bar{k}$. At variance, the RNN is capable to correctly discriminating the items provided as input entries only punctually, at $\bar{k}=5$. It loses any predictive power for $k>\bar{k}$.}
	\label{f:takememory}
\end{figure}

As mentioned earlier, a specific advantage of the RSN model is the ability to keep memory of the final state for $k>\bar{k}$. This is a byproduct of the fact that, for sufficiently large times,  the non-linear activation terms are virtually silenced and the update rule converges to a simple linear scheme. The attractors of the dynamics are by construction fixed points of the linear mapping, and this makes it possible to operate the RSN for any $k$  larger than the training horizon $\bar{k}$. As a benchmark model, we consider a standard Recurrent Neural Network (RNN) trained in direct space \cite{sherstinsky2020fundamentals,goldberg2017neural,medsker2001recurrent}. The RNN is conceived as a single transfer layer between two adjacent stacks made of $N=784$ nodes, iterated $k$ times (recognition is performed on the first $10$ nodes of the final layer).  The number of trainable parameters is thus $N \times N$, comparable to the number of parameters adjusted by the RSN model. In Figure \ref{f:takememory}, we compare the accuracy measured for the MNIST dataset, for both the RSN and the RNN trained upon completion of iteration $\bar{k}$. The accuracy recorded for the RSN (red symbols) converges rapidly and the achieved score is stably maintained for $k>\bar{k}$ (here $\bar{k}=5$). Conversely, the RNN (blue symbols) returns its largest accuracy (basically identically to that obtained with the RSN) only for $k=\bar{k}$. By taking just one step further (i.e. adding one additional layer to the RNN) is enough to lose predictive power.

\begin{figure}[H]
	\centering
	\includegraphics[width = 0.5\textwidth]{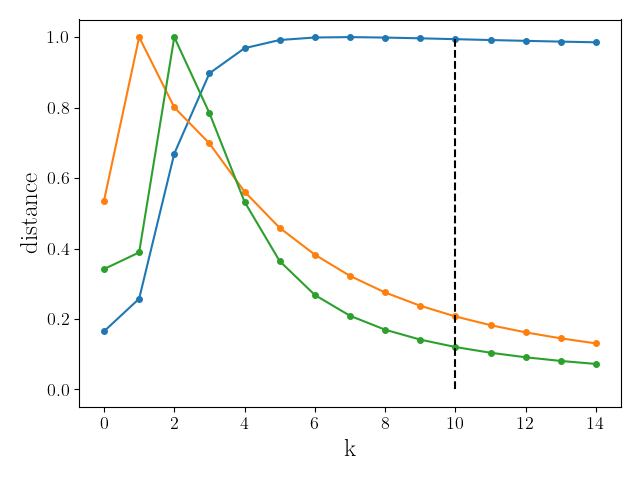}
	\caption{Euclidean distance (normalized to its maximum value) between the three sets of coefficients as described in the main body of the paper and respectively referred to fours and fives, against the iteration index $k$. The orange curve refers to ($10$) coefficients, associated to modes with small magnitude, as obtained after the training. The green curve is computed by considering the projections along ($10$) modes with eigenvalues bearing larger absolute values, though smaller than one.  The curve depicted in blue refers to the values of the coefficients of the $10$ eigen-directions relative to eigenvalues one, where classification is eventually performed. The peak travels horizontally suggesting that the information crawls from the transient towards the stationary modes. The fact that the orange curve seems more persistent that the green at larger $k$ is just a consequence of the imposed normalization. The vertical dashed line is set at $\bar{k}$.}
	\label{f:distanza}
\end{figure}

As also shown for the case of the simple model discussed in the preceding session, there is a progressive tendency to crystallize the final output in the eigen-directions, where recognition is eventually performed. The information steadily flow from the modes associated to the eigenvalues with smaller magnitude (the first to fade away as the iterations grow) to those characterized by eigenvalues with larger norms (more persistent than the former). This observation is made quantitative in Figure \ref{f:distanza}. Here, three sets of coefficients $c_i$ are identified. Each group clusters together different coefficients associated to eigen-directions relative to eigenvalues that approximately share the same magnitude (a set relative to small eigenvalues, a set relative to larger eigenvalues and the final set of eigenvalues equal to one, i.e., those associated to the eigen-directions where recognition takes place). We evaluate the three sets of coefficients for each image in the test set displaying a four and a five and compute the average distance (square norm) between each set of coefficients, against $k$, the iteration of the RSN. The coefficients stemming from the transient modes single out the differences between the analyzed samples, before converging to zero when the stationary attractors, inactive at first, get eventually approached 

In the next Section we will turn to considering a variant of the RSN which is constructed to yield sequential handling of different datasets, with a long term memory effect. To demonstrate our findings, and as a preliminary proof of concept, we will split MNIST into two distinct, though perfectly balanced, datasets, the first formed by digits from zero to four, and the other populated with the remaining elements, ranging from five to nine.

\section{Sequential learning: spectral quasi-orthogonality and the memory effects}

In this Section we will discuss a generalization of the RNS which allows to keep track, to some extent, of a learned task, while dealing with an independent session of training, on a distinct dataset. To elaborate along these lines, and with the sole aim of providing a preliminary proof of concept of the basic implementation, we shall split the MNIST into two distinct, though balanced datasets. The first will be composed by handwritten digits ranging from zero to four. The remaining images, displaying numbers from five to nine, constitute the second reservoir. We will then train the RSN to classify the images belonging to the first dataset. Then, the obtained RSN undergoes a second round of training focusing on the images that define the complementary dataset. By assuming  sets of quasi-orthogonal eigenvectors with associated memory kernels,  yields a fully coupled network, the backbone of the RSN, which is capable to efficiently handle novel tasks while preserving notion of past knowledge. This is at variance of conventional schemes, based on  standard deep learning architectures or RNN, which tend to eradicate former imprints by overwriting existing memory slots, as we shall hereafter demonstrate \cite{mccloskey1989catastrophic,lewandowsky1995catastrophic,kemker2018measuring}. 

\begin{figure}[H]
	\centering
	\includegraphics[width = 0.7\textwidth]{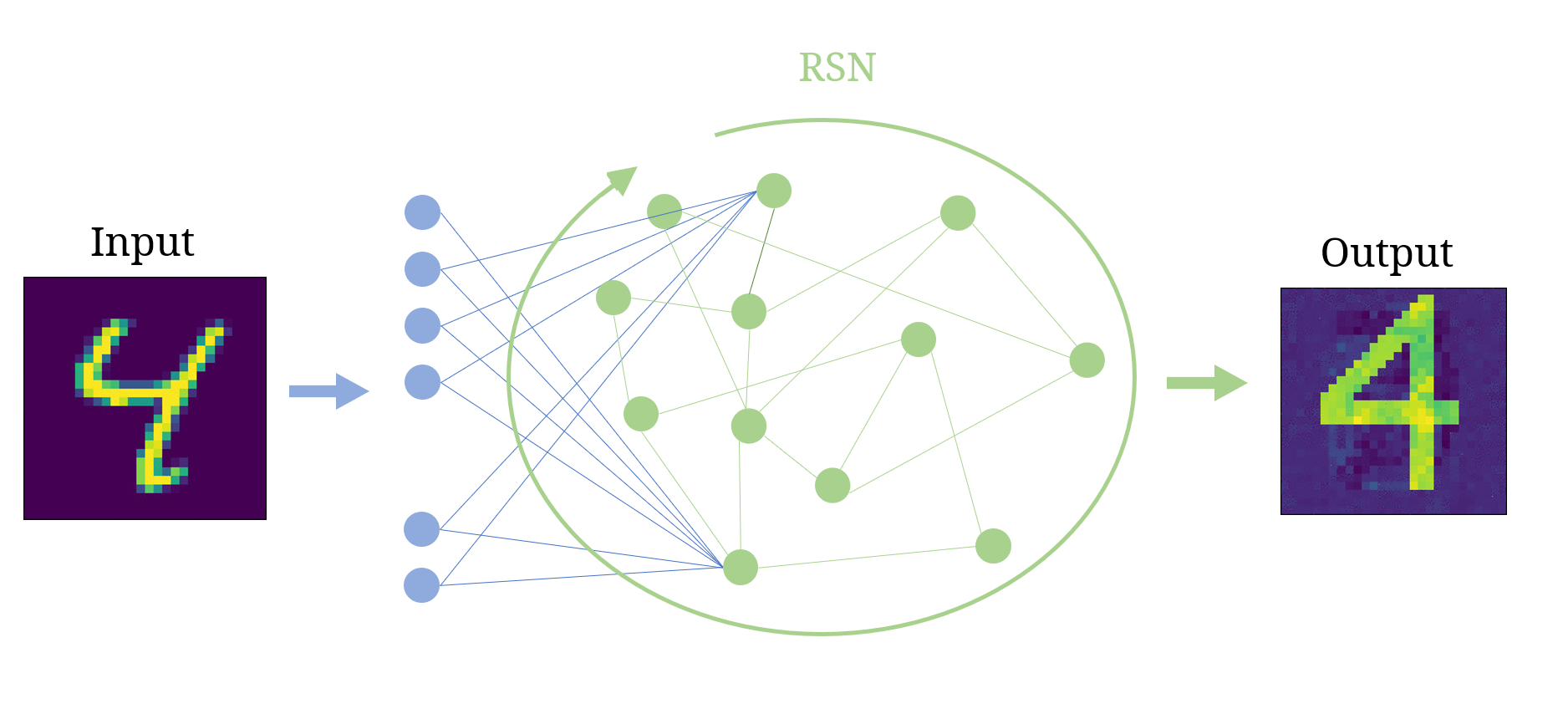}
	\caption{A schematic layout of the architecture employed to handle sequential learning. The information stemming from the image presented as an input are passed to the RSN, and therein iteratively elaborated until convergence to the deputed stationary stable attractor (here exemplified as a stylized version of the input number).}
	\label{f:schema}
\end{figure}

MNIST images are read as an input by a layer made of $N_0=28 \times 28$. This information is passed to the $N$ nodes of the RSN via an all-to-all linear transformation  encoded by a $N_0 \times N$ matrix ${\mathbf A_0}$, see Figure \ref{f:schema}. Here, ${\mathbf A_0}$ is fixed. As such, the entries of ${\mathbf A_0}$ do not take active part to the optimization process which is instead focused on the RSN component of the dynamics. Further, $N$ (assumed even, with no loss of generality) can be larger of smaller than $N_0$ without any limitation whatsoever. We then  postulate the following form for matrix ${\mathbf \Phi}$:

\begin{equation}
{\mathbf \Phi} = \begin{pmatrix}
  
  {\mathbf \Phi_{11}}
  
  & \rvline & \epsilon {\mathbf \Phi_{12}} \\
\hline
  \epsilon {\mathbf \Phi_{21}}  & \rvline &
   {\mathbf \Phi_{22}}
 \end{pmatrix}
\end{equation}

\begin{figure}[H]
	\centering
	\includegraphics[width = 0.6\textwidth]{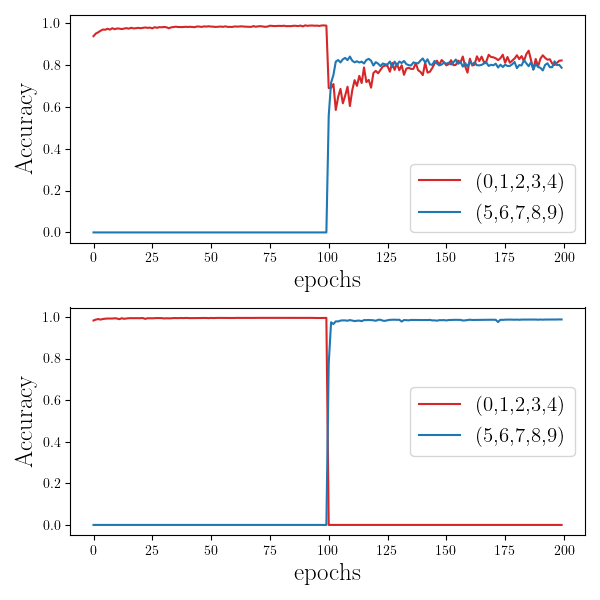}
	\caption{Top panel: accuracy against the epochs number for the RSN. The first $100$ epochs refer to the RSN confronted with the task of classifying the images of the dataset made of digits from zero to four. Then, the second range of epochs, refers to the RSN while learning to classify numbers from five to nine, after having completed the first stage of training. The accuracy drops but the RSN keeps still memory of the first task, while learning to cope with the second with an almost identical score of reported success. In this specific example, the elements of the off diagonal blocks ${\mathbf\Phi_{12}}$ and ${\mathbf\Phi_{21}}$ are kept fixed, during optimization. Lower panel: sequential learning is ineffective with usual RNN (and standard feedforward deep neural networks, data not shown), since any form of pre-installed knowledge gets washed out during a subsequent, independent, training stage. Here $\bar{k}=10$, $\epsilon=0.25$ and $N=1000$. }
	\label{f:sequential}
\end{figure}

The four blocks  ${\mathbf\Phi_{ij}}$, with $i,j=1,2$ have dimensions $N/2 \times N/2$, and comparable norms. The parameter $\epsilon$ sets the importance of the off-diagonal blocks as compared to those that define the block diagonal terms.  In the limiting case $\epsilon=0$ the matrix of the eigenvectors is block diagonal. The eigenvectors are hence organized into two distinct ensemble, mutually orthogonal and the corresponding network split into two disconnected parts. When  $\epsilon \ne 0$ instead the two subparts of the ensuing network get mutually entangled and virtually indistinguishable for a sufficiently large magnitude of the coupling parameter $\epsilon$. For $\epsilon \ne 0$, though relatively small as we shall assume in the following, the eigenvectors form two quasi-orthogonal blocks.  Focus now on the diagonal matrix of the eigenvalues. These are also split into two groups of identical cardinality, which will be eventually structured as follows $\left(1, 1, 1, 1, 1, \lambda_6, ...,  \lambda_{N/2} \right)$ and $\left(1, 1, 1, 1, 1, \lambda_{N/2+6}, ...,  \lambda_{N} \right)$. Trivial eigenvalues are associated to specific eigen-directions, the attractors of the RSN, which stay put across optimization. In practice, each eigenvalue  equal to unit points to a specific memory slot which can be filled and, at least partially, preserved, across multiple learning stages. Starting from this setting we proceed as follows:

\begin{itemize}
\item We set at first to zero the first five eigenvalues belonging to the second group, as identified above. In doing so, we seek at protecting a specific set of memory slots, which should not be contaminated during the first round of training
\item We then train the RSN to recognize and correctly classify the first reservoir made of handwritten digits from zero to four, as outlined above. During this operation, the optimization acts on $\lambda_6 ...  \lambda_{N/2}$ and on (the full set or a limited sub-portion of) the entries of the eigenvectors associated to these latter eigenvalues. Here, $\epsilon \ne 0$, which in turn implies that by modulating the entries of the eigenvectors belonging to the first of the two sets, yields an indirect signature on all the inter-nodes weights in direct space. At the end of the optimization, the RSN is capable to correctly classifying analogous images belonging to the test set.
\item We  then turn to the second round of training by providing to the above RSN (namely, the RSN that has been trained to cope with the first dataset) the elements belonging to the second reservoir of images, those depicting digits ranging from five to nine. The second set of memory slots is  turned on, by setting to unit the eigenvalues initialized to be zero: the corresponding eigenvectors define the asymptotic attractors where the information is eventually stored. The eigenvalues that identify the target eigen-directions from the preceding training are instead set to zero.   
\end{itemize}
After completion of the optimization, one can check the performance of the RSN, which has been trained across two successive stages, referred to two distinct datasets. To this end we turn on all possible memory slots (trained as follows the above, two steps, procedure): in practice we set to one the eigenvalues relative to the (10) eigen-directions where information is asymptotically conveyed. In Figure \ref{f:sequential} (top panel), the performance of the RSN, as measured by the reported accuracy, is tested against the epochs of the optimization scheme. The optimization is carried out by assuming $\bar{k}=10$ in the RSN, and assuming 100 of epochs for each of the two nested stages of learning. Already after a few epochs the RSN returns a very high accuracy against images of the test set which display digits ranging from zero to four. When the RSN gets also  trained on the complementary reservoir of handwritten digits, as follows the sequential scheme highlighted above, it quickly manages to handle the novel task with an adequate success rate, while, at the same time, manifesting a relatively modest drop in performance as referred to the former. Notice that the images are supplied as an input with no extra markings, or alert flags, to point to the relevant group of attractors.  To grasp the interest of the proposed scheme we report in Figure \ref{f:sequential} the results obtained for a RNN with a number of layers equal to $\bar{k}+1$. As immediately confirmed by visual inspection, any knowledge coming from the first round of training is - almost instantaneously - lost, when the network becomes acquainted with the second  task. Similar conclusions (data not shown) are obtained when dealing with a deep neural network, with a standard feedforward architecture \cite{mccloskey1989catastrophic,lewandowsky1995catastrophic,kemker2018measuring}. Summing up, working with a quasi-orthogonal basis, with a set of (almost) mutually exclusive blocks equal to the number of tasks to be eventually handled, yields a RSN which can be sequentially trained, while keeping memory of the previous training sessions. A drop in the recorded accuracy is however found which could be possibly mitigated for increasing RSN size and/or addressing ad hoc solutions that require further investigations, beyond the scope of a mere proof of concept. It is also remarkable that the accuracy displayed against the first dataset, and after the initial sudden jump that follows the second training round, ramps again, epoch after epoch, to align to that refereed to the second dataset.

\section{Conclusions}

In this paper we have introduced and tested a novel approach to automated learning, which is rooted in reciprocal space and exploits foundational elements of the theory of discrete dynamical systems. The information under scrutiny is read by a collection of nodes, typically (but not necessarily) the pixels of the image provided as an entry, and further processed by the very same nodes, as follows an iterative update scheme which alternates linear mapping and non-linear filters.
Depending on the characteristics of the signal provided as an input, the ensuing dynamics is steered towards different asymptotic attractors, where recognition takes eventually place. The convergence to the asymptotic state is stable by construction, and the alignment along the selected direction is guaranteed also when pushing the iterations beyond the limited horizon of the optimization. We have referred to the proposed methodology as to Recurrent Spectral Network RSN, to signify the dynamical nature of the process which is formulated in reciprocal domain.  

Neural networks are sometimes called black boxes because it is not immediate to understand how or why they work as well as they do. At variance, the operational mode of a RSN is absolutely transparent and, as such, it could help unveiling the blanked of mystery that surrounds machine learning applications. Indeed, the RSN associates different classes of items to distinct attractors of an underlying discrete dynamical systems. Learning to classify within the RSN amounts to correctly shape the separatrix between adjacent attraction basins, a task that is accomplished by embedding the data to be examined into a space of  sufficiently high dimensionality. Non-linearities act over a transient and fades eventually away, when the non trivial classification problem has been de facto turned into a linear one. 

A  variant of the RSN has been also considered which accounts for quasi-orthogonal eigen-directions to carry out a sequential handling of different datasets. In practice, a RSN can be assembled which keeps memory of an initial task, while being subject to another session of training on an independent dataset. 

Several directions for further investigations can be outlined. One interesting possibility is to modify the loss function by forcing the contribution at iteration $k$ to be smaller than that at iteration $k+1$. Preliminary checks shows that the RSN tunes self-consistently its convergence rate, which is hence not a priori imposed as it is here done. It is also tempting to speculate that proceeding along these lines, one could eventually generate a RSN which is capable of improving its accuracy score by iterating further beyond the specific window of training. Another possibility is to introduce apposite frustration mechanisms, which tend to disfavour the accidental convergence towards directions that have been already exploited, when operating with the sequential learning protocol. Also, it would be extremely important to devise other possible strategies, alternative to the one here employed, to structure the eigenvectors matrix for multiple datasets handling.        

 \bibliographystyle{apsrev4-1}
\bibliography{references}

\end{document}